\documentclass[12pt,preprint]{aastex}

\shorttitle{Optical and X-ray variations in NGC~5548}
\shortauthors{Uttley et al.}

\begin{document}


\title{Correlated long-term optical and X-ray variations in NGC~5548}


\author{Philip Uttley\altaffilmark{1}, Rick Edelson\altaffilmark{2},
Ian M$^{\rm c}$Hardy\altaffilmark{1}, Bradley M. Peterson\altaffilmark{3}
\and Alex Markowitz\altaffilmark{2}}

\altaffiltext{1}{Department of Physics and Astronomy, University of
Southampton, Southampton SO17 1BJ, UK.  pu,imh@astro.soton.ac.uk}
\altaffiltext{2}{Astronomy Department, University of California, Los
Angeles, CA 90095-1562.  rae,agm@astro.ucla.edu}
\altaffiltext{3}{Department of Astronomy, The Ohio State University, 140
West 18th Avenue, Columbus, OH 43210-1173. 
peterson@astronomy.ohio-state.edu}


\begin{abstract}
We combine the long-term optical light curve of the Seyfert~1 galaxy
NGC~5548 with the X-ray light curve measured by the {\it Rossi X-ray
Timing Explorer} over 6 years, to determine the relationship between the optical
and X-ray continua.  The X-ray light curve is strongly correlated with
the optical light curve on long ($\sim$year) time-scales.
The amplitude of the long-term optical variability in NGC~5548 is larger
than that of the X-ray variability (after accounting for the host galaxy
contribution), implying that X-ray reprocessing is not the main source of
the optical/X-ray correlation.  The correlated X-ray and optical
variations in NGC~5548 may be caused by instabilities in
the inner part of the accretion flow, where both the X-ray and optical
emission regions may be located.  
\end{abstract}




\section{Introduction}
It is commonly supposed that the X-ray and optical emission in
Seyfert~1 galaxies
is produced by different, but possibly related physical processes.  The X-rays are thought to
be produced by Comptonization 
of lower energy `seed photons' by a hot optically thin corona, while
optical photons arise as thermal
emission from the cooler, optically thick accretion disk.  The relationship
between flux variations in the optical and X-ray band
can be used to probe the relationship between the X-ray
and optical emitting regions.  Optical/X-ray
correlations might be produced if the emitting regions are somehow
connected, e.g. if variable optical emission is
caused by X-rays heating the disk, or if the optical photons are the seed
photons in the X-ray Comptonization process.   On the other hand, if the emitting
regions are physically separated, we might expect to see no
correlation between optical and X-ray variations.

The current picture regarding correlated optical/X-ray variability in
AGN is ambiguous.  In early work on the optical/X-ray
correlation in NGC~4051, Done et al. (1990) showed there was very
little optical variability ($<1$\%) on time-scales of days, and no apparent
correlation with the much larger amplitude X-ray variability, although
a possible short-term UV/X-ray correlation in this source has been recently reported by
\citet{mas02}.  Also, using
long-term optical and X-ray monitoring observations of NGC~4051,
\citet{pet00} suggested that the optical and X-ray light curves were
correlated on time-scales of months-years.
Early, sparsely-sampled monitoring
suggested a correlation between ultraviolet and X-ray variations in NGC~5548
\citep{cla92} and NGC~4151 \citep{ede96}.  A
possible 100~day X-ray-to-optical lag in NGC~3516, based on 
$\sim1.5$~yr duration optical and X-ray monitoring lightcurves
\citep{mao00}, was not confirmed with further
monitoring over an additional 3.5~yr \citep{mao02}.  An intensive 3 day
observation of NGC~3516 found no correlation between optical ({\it HST})
and X-ray variations \citep{ede00}.  Intensive X-ray and {\it
IUE} monitoring
of NGC~7469 showed no correlation between X-ray and UV flux
\citep{nan98}, but there does appear to be a correlation between UV flux and X-ray
energy spectral index \citep{nan00}.  
In Ark~564, a possible weak optical/X-ray correlation (based on a
single flaring event) was observed \citep{she01}.  In general, the amplitude of
optical variations in AGN appears to be less than that of X-ray
variations, especially on short time-scales \citep{pet00,she01,mao02}.

In this Letter, we re-examine the optical-X-ray correlation in NGC~5548,
combining 6 years of {\it Rossi X-ray Timing Explorer} ({\it RXTE})
monitoring observations with the
corresponding optical light curve from the International AGN Watch
monitoring campaign.

\section{Observations and data reduction}
For 13 years, from late 1988 until the end of 2001, NGC~5548 was monitored
in the optical band by the International AGN Watch
in order to reverberation map its broad line region \citep{kor95,pet02}.
Since 1996 May, NGC~5548 has been
monitored in X-rays by the {\it Rossi X-ray Timing Explorer} ({\it RXTE}),
using sampling intervals ranging from 8 times daily to every two weeks,
primarily to measure its X-ray variability power spectrum
\citep{utt02,mar02}.

We obtained the latest optical 5100\AA\
continuum flux light curve from the International AGN Watch public
archive\footnote{available at URL
http://www.astronomy.ohio-state.edu/$\sim$agnwatch/.}.  Note that all
light curves from the International AGN Watch are corrected for
aperture effects which might otherwise lead to spurious variability, e.g. due to
varying host galaxy starlight contamination (see Peterson et al. 2002
for calibration details).  We used data from the
PCA instrument on board {\it RXTE}, reduced in the standard manner and with
model background data generated using the latest background models and the
{\sc ftools v5.0} software (see Lamer, Uttley \& M$^{\rm c}$Hardy
2000, for details).  Changes in 
the instrument gain throughout the lifetime of the {\it RXTE}
mission mean that it is not straightforward to simply compare instrument
photon count rates in a given band.  Therefore,
for each {\it RXTE} monitoring observation
(which are snapshots of length $\sim300-1000$~s), we extracted a spectrum
and fitted a simple power-law plus 6.4~keV Gaussian model to determine the flux
in the 2--10~keV band.   Note that for the purposes of measuring the
correlation between the light curves, neither the optical or X-ray
light curves are corrected for the
contaminating contributions due to host-galaxy starlight
($3.4\times10^{-15}$~erg~s$^{-1}$~cm$^{-2}$~\AA$^{-1}$ at 5100\AA\,
accounting for $\sim34$\% of the mean total 5100\AA\ flux,
Romanishin et al. 1995) or the presence of a BL Lac object in the {\it RXTE} 
PCA field of view ($\sim10$\% of observed flux or 
$\sim6\times10^{-12}$~erg~s$^{-1}$~cm$^{-2}$, extrapolated from data in
Chiang et al. 2000).  Clearly the constant contribution from
starlight will not affect any correlation, while the additional X-ray 
variability due to the BL Lac will act to reduce any correlation that
we see, rather than produce a spurious correlation.

\section{Light Curves and Cross-correlation Function}
\begin{figure}
\epsscale{1.0}
\plotone{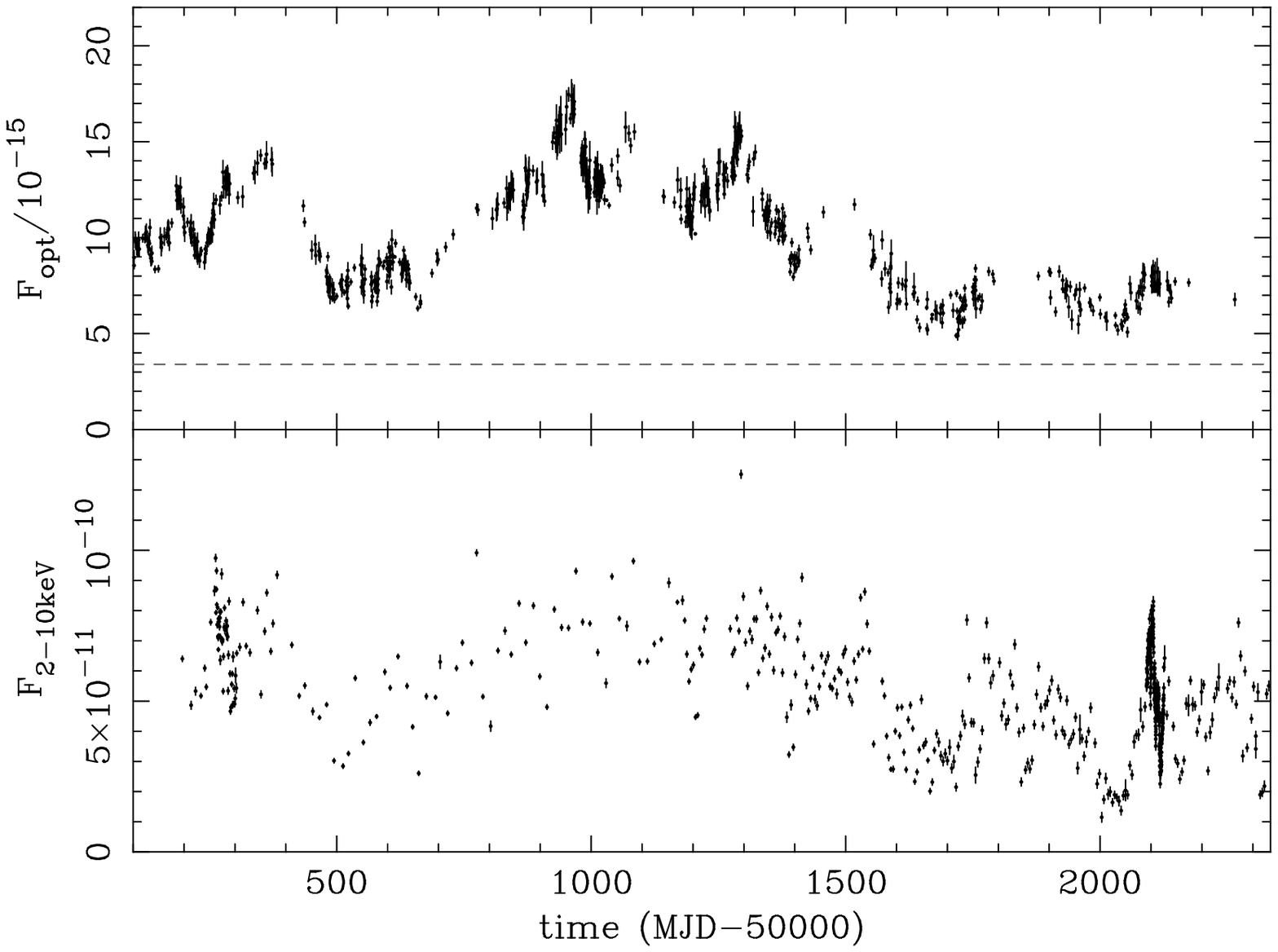}
\caption{Long-term optical 5100\AA\ (top) and X-ray 2-10~keV (bottom) 
light curves of
NGC~5548, in units of erg~s$^{-1}$~cm$^{-2}$~\AA$^{-1}$ and
erg~s$^{-1}$~cm$^{-2}$ respectively.  The
constant level of contamination due to host-galaxy starlight (see Section~2) has not been
subtracted from the optical lightcurve, but is shown for information
purposes as a dashed line.} \label{rawlcs}
\end{figure}

\begin{figure}
\epsscale{1.0}
\plotone{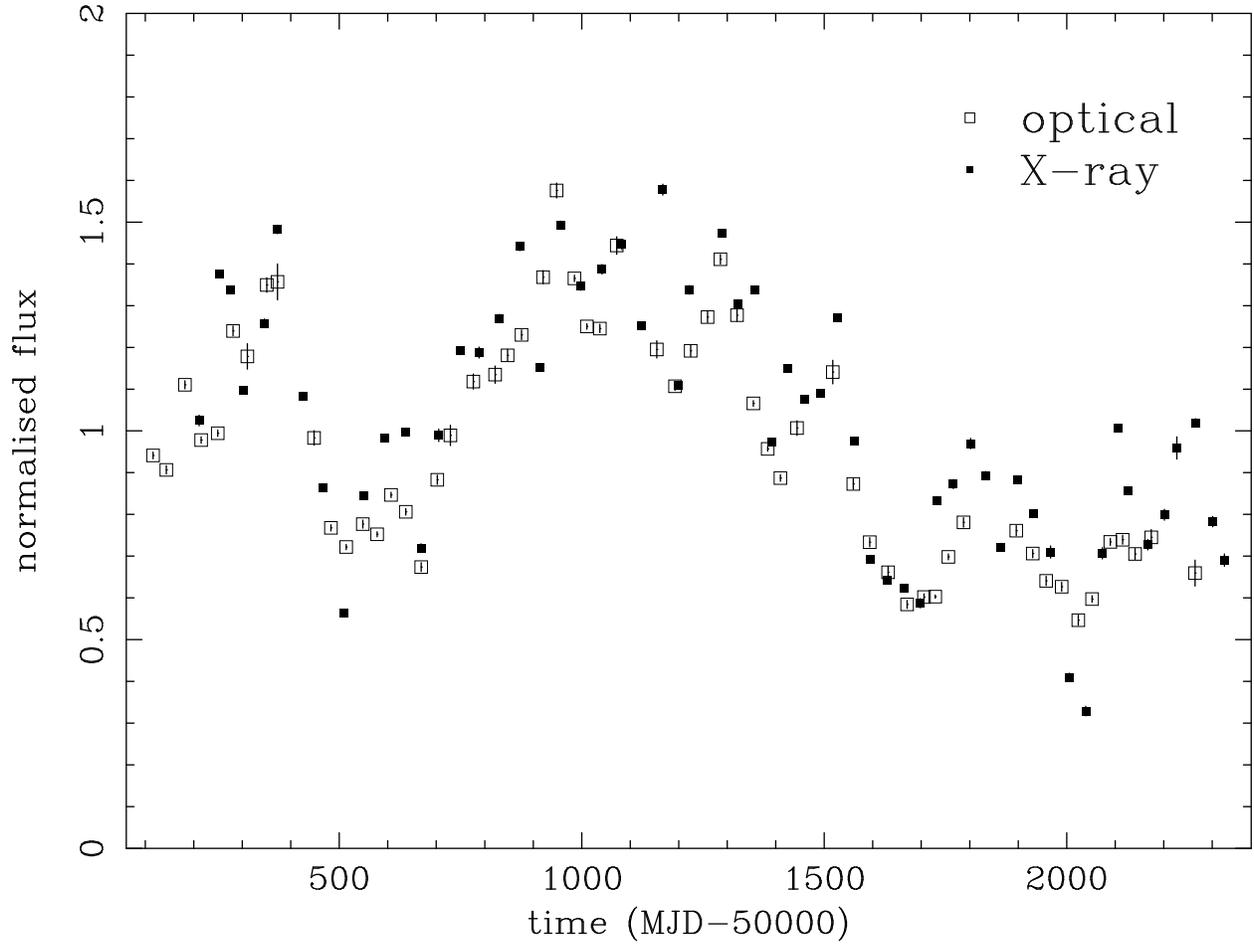}
\caption{Binned-up long-term optical 5100\AA\ (open squares) and X-ray 2-10~keV
(filled squares) light curves of NGC~5548.  Bin time is 30~d and the
light curves have been renormalized by their respective means.  The
constant level of contamination due to host-galaxy starlight (see Section~2) has not been
subtracted from the optical lightcurve.} \label{binlcs}
\end{figure}

The optical 5100\AA\ and X-ray 2-10~keV light curves are shown in
Figure~\ref{rawlcs}.  On short time-scales (days), the X-rays show more rapid
variations than the optical light curve.  But despite this difference, and the
relatively sparse X-ray sampling throughout the first half of the {\it RXTE}
campaign, an apparent correlation of the long-time-scale X-ray and optical
variations can be discerned. 
 To show this correlation more clearly,  we
smooth out the short time-scale variations
by binning both light curves into 30~d bins, and plot the resulting
light curves, renormalized by their respective means, in Figure~\ref{binlcs}.
The correlation extends not just to the general rising then falling trend of
the light curve, but also to variations on shorter time-scales.

\begin{figure}
\epsscale{1.0}
\plotone{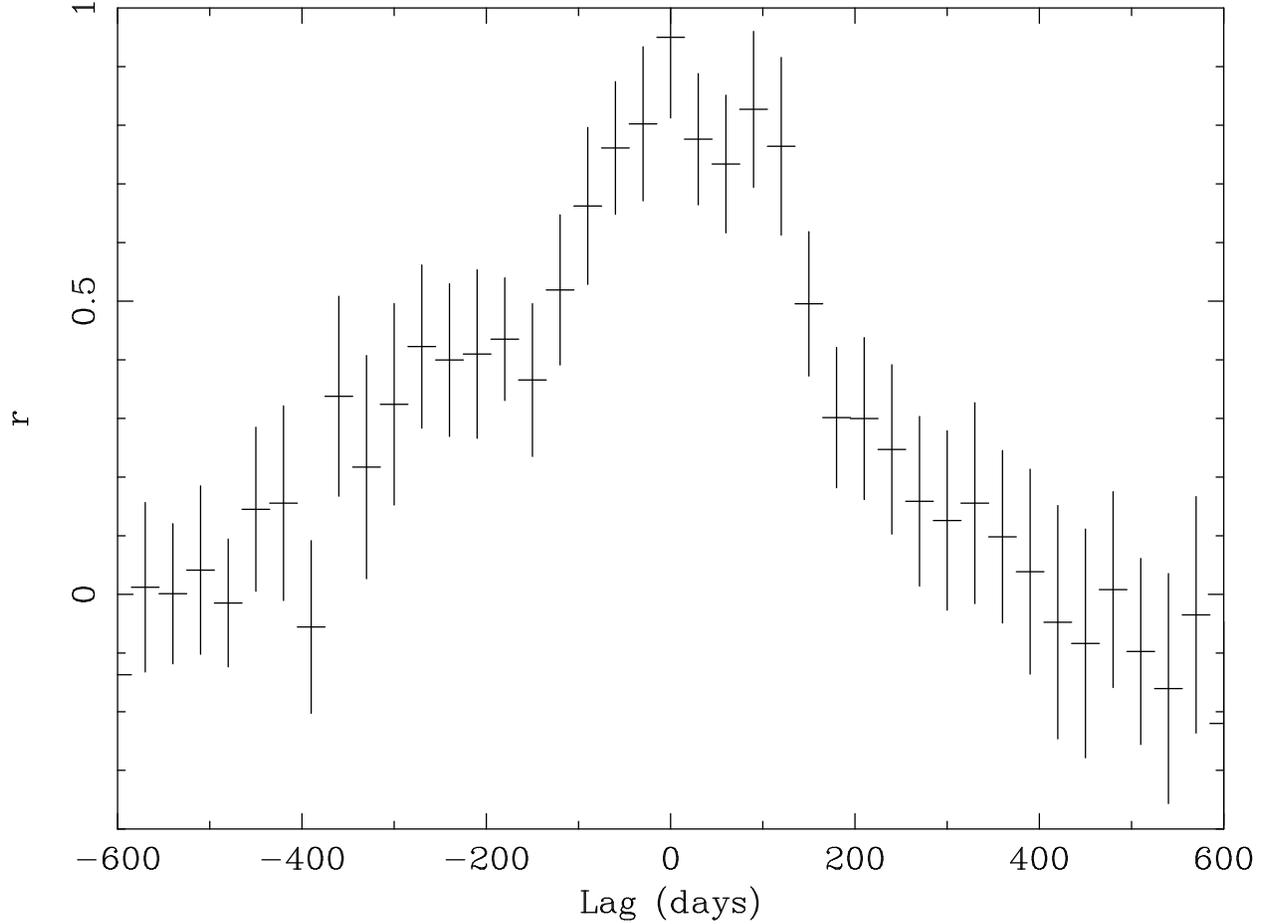}
\caption{Cross-correlation function of optical and X-ray light curves. 
Note: the error bars on the correlation are shown for convention purposes
only.  For red-noise light curves, standard errors in CCFs are correlated and do
not have the usual, formal meaning.} \label{ccfplot}
\end{figure}

We confirm the apparent strong correlation between the X-ray and
optical light curves by measuring the cross-correlation
function (CCF) of the 30-d binned light curves (see Figure~\ref{ccfplot}),
using the Discrete Correlation Function (DCF) method of Edelson \& Krolik
(1998).  The peak value, $r_{\rm max}$ of the 30-d binned lightcurve's CCF, at zero
lag, is $r_{\rm max}=0.95$, compared to
$r_{\rm max}=0.85$ observed in the CCF of the unbinned lightcurves.  Using the
Flux Randomization/Random Subset Selection (FR/RSS) method
of \citet{pet98}, with both binned and unbinned
 light curves, we find the lag of the CCF peak is consistent with
$0\pm15$~d (1-$\sigma$ error).

Unfortunately, we cannot assess the significance of this correlation
directly from $r_{\rm max}$, since that assumes that individual data points in each
light curve are uncorrelated with adjacent points, when in fact they
are correlated, `red-noise' data.  Thus, the effective number of data
points in the correlation, and hence its significance, is reduced by
an amount which depends on the sampling pattern and power-spectral
shape of both light curves, and cannot be determined analytically.  We therefore test the
significance of the correlation using Monte-Carlo simulations of uncorrelated
red-noise light curves.  We used
the method of Timmer \& K\"{o}nig (1995) to simulate $10^{4}$ pairs of
continuous red-noise light curves (one for each band) of time
resolution 0.1~d, assuming appropriate power spectral shapes and
normalization for each band\footnote{We assumed a broken power-law
shape power spectrum, with high-frequency
slopes of -1.6 and -2.5 in X-ray and optical bands
respectively and identical slopes of -1 in both bands
below a break frequency of $\nu<0.01$~d$^{-1}$,
as implied by scaling to the power spectra of black hole X-ray binaries
\citep{utt02} and assuming a $\sim10^{8}$~M$_{\odot}$ black-hole mass estimated from 
reverberation mapping \citep{wan99}.  The high-frequency
power-spectral slopes and assumed normalization are the best-fits to the data assuming this
break frequency and low-frequency shape (see \citet{utt02} for details of the power-spectral
fitting procedure).}.
Each simulated light curve was resampled to the sampling pattern of the
corresponding observed light curve and observational noise added, in the form
of a Gaussian deviate with mean zero and standard deviation equal to the
average error of the observed light curve.  We then rebinned the
simulated,
resampled light curves to the same 30~d bins as used in our real analysis,
measured the CCF using the DCF method, and counted the number of
simulated CCFs containing peak values $r_{\rm max}>0.95$.

When searching for peaks in the plausible lag range of 
$\pm200$~d, peak CCF values with $r_{\rm max}>0.95$ were
observed in zero out of $10^{4}$ simulations.  We conclude
that the observed long-term correlation between the optical continuum and
X-ray bands in NGC~5548 is significant at better than 99.99\% confidence.

\section{Amplitudes of variability} \label{varamp}
As is apparent in Figure~\ref{binlcs}, the amplitude of variations in both
bands is comparable on long-time-scales ($>$months).  In fact, when one
subtracts the estimated contaminating contributions in the X-ray and optical
bands from BL~Lac object and host-galaxy starlight respectively, one finds that
the optical 30~d binned light curve is more variable than the 30~d
binned X-ray light curve (fractional rms variability
$F_{var}=43\pm0.2\%$ versus $31\pm0.1\%$
respectively, where errors are statistical errors due to observational noise
only).  This result is particularly robust, since even
assuming the 68\% confidence lower limit on
optical contamination from the host galaxy
($2.83\times10^{-15}$~erg~s$^{-1}$~cm$^{-2}$~\AA$^{-1}$, Romanishin et
al. 1995 and references therein), any constant, contaminating
component in the X-ray lightcurve would have to contribute $\sim29$\% of
the observed X-ray flux, simply for the underlying X-ray variability
amplitude to match that in the optical band.  Note that the BL~Lac
contribution to the X-ray lightcurve
is likely to be only 10\% of the total flux, and is not constant \citep{chi00},
so it may in fact contribute to the observed X-ray variability, rather
than simply reducing its fractional amplitude.
The unbinned lightcurves (Figure~\ref{rawlcs}) show that, on short time-scales ($<30$~d)
the amplitude of X-ray variations is
larger than the amplitude of optical variations, contributing
$\sim10$\% of the total rms variability in the X-ray band, as opposed to $\sim2$\%
contributed by short-time-scale variations in the optical band.  This result is also implied by
the X-ray power spectrum being flatter than the optical power
spectrum at high frequencies (e.g. Uttley et al. 2002, Collier \&
Peterson 2001).

\section{Discussion}
We have shown that the long-term X-ray and optical continuum 
light curves of NGC~5548 are both highly variable and 
highly correlated.  This result contrasts with the situation so 
far observed in other AGN, where X-ray and optical/UV 
lightcurves are either not very well correlated 
(e.g. Nandra et al. 1998, Shemmer et al. 2001, Maoz et al. 2002), or a
long-term optical/X-ray correlation is seen but
the optical lightcurves are only weakly variable compared to the X-ray
lightcurves (NGC~4051, Peterson et al. 2000).  Note also that although the absolute
time-scales (months-years) for correlated optical/X-ray variability
in NGC~5548 are similar to those observed in NGC~4051, the relative
time-scales (in terms of characteristic time-scales of the system) are
much shorter, since NGC~5548 has a factor $\sim100$ larger black hole
mass than NGC~4051 \citep{wan99}.

It is possible that the X-ray/optical production mechanisms in 
NGC~5548 are physically different to those in other AGN.  Perhaps
the optical variability in NGC~5548 is driven by X-ray reprocessing,
whereas other processes are dominant in other AGN.
 However, the large amplitude of optical variability poses problems for
the reprocessing model.  One would expect that any reprocessed emission would
exist {\it in addition to} a non-reprocessed component due to intrinsic 
thermal emission from the disk, so that the amplitude of long-term optical
variability should be less than the amplitude of X-ray variations.  In
contrast, we see that on long time-scales,
the optical variations in NGC~5548 are larger than the X-ray
variations.  The problem for reprocessing models is further compounded
by the spectral variability in the optical and X-ray bands.
Specifically, the X-ray spectrum softens as the 2-10~keV flux
increases \citep{chi00}, so that the
broadband X-ray luminosity probably varies less than the 2-10~keV variations we
observe here.  At the same time, the UV continuum is strongly correlated with and
even more variable than the optical (with $F_{\rm
UV}\propto F_{\rm opt}^{1.8}$, Peterson et al. 2002),
so that the total optical/UV luminosity is even more variable than the
5100\AA\ continuum lightcurve we show here.  In defence of
reprocessing models, one could invoke a large variable component
available for reprocessing in the hidden
EUV band which may itself be Comptonized thermal emission
\citep{mag98}.  However, the response of the
H$\beta$ emission line to changes in the optical/UV continuum is consistent with
photoionization calculations, which assume the ionizing (EUV ) continuum
simply tracks the UV continuum \citep{gil02}.  Therefore it is
unlikely that the EUV continuum is much more variable than the UV.

Assuming the standard disk-corona model for the continuum emission in
AGN, it is almost certain that at least some component of the optical
variability is associated with X-ray reprocessing.
However, an additional component is probably required to
explain the high amplitude of optical
variability in NGC~5548.  The simplest possibility is that both X-ray
and optical variations are tracking some other, more fundamental
variations in the accretion flow, for example thermal instabilities in
the inner disk \citep{tre88}.
Provided that the X-ray and optical emission regions are close
together, such variations could lead to correlated X-ray variability,
e.g. through optical seed photon variations modulating X-rays
directly (as suggested by the correlated X-ray spectral and UV 
variability seen in NGC~7469, Nandra et al. 2000), or
through simple viscous heating
in the disk correlating with coronal heating by magnetic reconnection.
 The additional short-term X-ray variability may be caused by additional
variability processes in the X-ray emitting region (e.g. magnetic
reconnection) or by the X-rays being relatively more concentrated towards the
centre of the accretion flow, where variability time-scales are shorter.

It is apparent
that, in this picture, the relation between optical and X-ray variability in
an AGN should depend on the location of the optical emitting region
relative to the X-ray emitting region.  Comparisons of the X-ray
variability power spectra of AGN and black
hole X-ray binaries are consistent with the interpretation that, in units of Schwarzschild radii
($R_{\rm S}$), the size of the
X-ray emitting region is independent of black hole mass
\citep{utt02,mar02}.  However, since the disk temperature is dependent
on the black hole mass and accretion rate, with AGN with lower mass
black holes or higher accretion rates having higher disk temperatures,
the region of peak optical emission should occur at larger radii
(in $R_{\rm S}$ units) in these AGN.  It follows that in AGN with
lower black hole mass and/or higher accretion rate,
the optical emitting region exists further from the X-ray
emitting region, where variability time-scales are longer than the
inner regions  and/or the
disk may be more stable, so that the optical emission varies less than the X-ray emission
and may also respond to variations which the X-ray emitting region does not 
even see (and vice versa).

We illustrate this simple physical picture by considering the emission from a standard
thin disk.  At each small increment of disk radius, we assume local black body emission
with a temperature determined assuming the radial
temperature dependence of a thin disk \citep{fra92}:
\begin{equation}
T=3.53\times10^{7}\:\left(\frac{\dot{m}}{\eta M_{\rm BH} R^{3}}\right)^\frac{1}{4}
\:\:\:\:\:\:{\rm K}
\end{equation}
where $\dot{m}$ is the specific accretion rate, i.e. mass accretion rate
expressed as a fraction of the Eddington rate, $\eta$ is the
efficiency of accretion (here we assume the canonical value of $\eta=6$\% for
accretion on to a non-rotating black hole, e.g. Frank et al. 1992), $M_{\rm BH}$ is the black
hole mass in solar mass units, and $R$ is the radius in terms of Schwarzschild
radii.  By integrating
the derived flux at 5100\AA\ over a range of radii, we determine where
most of the 5100\AA\ emission originates.  Assuming a black hole mass in 
NGC~5548 of $\sim10^{8}$~M$_{\odot}$ (from reverberation mapping,
Wandel et al. 1999), and accretion rate
$\dot{m}\sim0.01$ \citep{wan99}, we find that 50\% of the 5100\AA\ emission is
contained within 42~$R_{\rm S}$ of the central black hole.  Given the
relatively rapid X-ray variability, it is likely that most of the
X-ray emission originates within this radius.  Assuming a
viscosity parameter $\alpha=0.1$, the corresponding thermal time-scale
at this radius is $\sim280$~d \citep{tre88}, consistent with the
timescales of variability we observe here.  By
contrast, in NGC~4051 if we assume $M_{\rm BH}\sim10^{6}$~M$_{\odot}$
and $\dot{m}\sim0.1$ \citep{wan99}, we find
that only 4\% of the 5100\AA\ continuum flux is emitted within
$R=42$~$R_{\rm S}$.  Similarly, in NGC~3516 
($M_{\rm BH}\sim2\times10^{7}$~M$_{\odot}$, Onken et al. 2002,  and
$\dot{m}\sim0.1$), only 14\% of the 5100\AA\ emission originates within
$R=42$~$R_{\rm S}$.  The precise numbers we
quote here should be taken very lightly, since the optical continuum of
NGC~5548 likely consists of a Comptonized component, in addition to
the black body emission we assume here \citep{mag98}.  Furthermore,
the thin disk model we use is almost certainly too naive, as it does
not account for, e.g. dissipation of accretion power in a corona
\citep{sve94}, or the possibility that the inner disks of the higher
$\dot{m}$  AGN may in fact be geometrically thick.  However, regardless of the
details of the model, it is reasonable to assume
that high mass, low $\dot{m}$ AGN such as NGC~5548 should have lower
temperature disks and hence more centrally-concentrated optical emitting
regions (i.e. closer to the X-ray emitting region)
than low mass, high $\dot{m}$ AGN.  This difference may help to
explain the confusing range of optical/X-ray correlations seen in
Seyfert galaxies so far.



\acknowledgments
RE and AM acknowledge support from NASA grant NAG 5-9023.  We thank
the anonymous referee for many helpful comments.





\clearpage

\clearpage

\clearpage

\end{document}